\documentclass[twocolumn,showpacs,showkeys]{revtex4}
\usepackage{graphicx}
\usepackage{bm}
\usepackage{color}
\usepackage{amsmath}
\usepackage{natbib}

\begin{document}

\title{Dipolar-drift and collective instabilities of skyrmions in crossed nonuniform electric and magnetic fields in a chiral magnetic insulator}

\author{Pavel A. Andreev}
\email{andreevpa@physics.msu.ru}
\affiliation{Faculty of physics, Lomonosov Moscow State University, Moscow, Russian Federation, 119991.}
\author{Mariya Iv. Trukhanova}
\email{mar-tiv@yandex.ru}
\affiliation{Faculty of physics, Lomonosov Moscow State University, Moscow, Russian Federation, 119991.}

 \date{\today}

\begin{abstract}
Skyrmions dynamics in the approximation of point-particle skyrmions is considered in the multiferroic insulators,
where the  spiral magnetic structure is accompanied by finite electric dipole moment.
The skyrmion motion is studied in the gradients of perpendicular magnetic and electric fields.
The nonuniform magnetic field with a fixed gradient magnitude leads to the spiral motion,
where skyrmion circulates in the plane perpendicular to the direction of magnetic field.
The frequency of circulation is fixed by the derivative of the external magnetic field and the magnitude of the dipole moment.
Application of additional electric field
which perpendicular to the magnetic field,
but their gradients have same direction leads to the drift motion of skyrmions resulting in the magneto-dipolar Hall effect.
Hydrodynamics of the gas of interacting skyrmions in crossed nonuniform electric and magnetic fields
shows the new type of hydrodynamical instability caused by the drift motion.
Three regimes of instability are described.
\end{abstract}




\maketitle


The concept of skyrmions appeared in the nuclear physics.
However, years later it become important in the condensed matter physics \cite{Muhlbauer Sc 09, X.Z.Yu Nat Mat 11, Heinze Nat P 11, Everschor PRB 12}
(for instance for the chiral cubic magnets $MnSi$ \cite{Muhlbauer Sc 09}),
including the field of quantum gases \cite{Herbut PRL 06, Kawakami PRL 12, Liu NJP 17, Tiurev NJP 18}.
Skyrmions are topologically nontrivial nonlinear structures formed of the magnetization vector field.
Their study is focused on development of compact memory cells for storing information \cite{Iwasaki Nat Nano 13}.
Skyrmions can be considered as a particle-like spin textures in thin films and bulk materials with broken inversion symmetry
and strong spin-orbit coupling under weak applied magnetic fields
\cite{Robler Nat 06, Fert Nat Rev Mat 17, Jiang Sc 15, Buttner Sc Rep 18, Psaroudaki PRX 17}.
The electric currents \cite{Jonietz Sc 10, Schulz Nat Ph 12, YH Liu CPL B 15}
and heat currents \cite{Mochizuki Nat Mat 14} are used for the manipulation of the motion of skyrmions.

The tasks of manipulating skyrmions and controlling their position remain important tasks and require finding of new mechanisms of manipulation.
To this end, a single skyrmion in a magnetic materials can be described as a particle-like object in certain regimes.
The equation of motion for skyrmions is derived in Refs. \cite{Brown PRB 18, Nagaosa Nat Nano 13} using Thiele's approach,
where the skyrmions are represented as rigid point-like particles.
The particle-based simulations of skyrmions is used to examine the static and driven collective phases of skyrmions,
interacting with random quenched disorder \cite{Reichhardt PRL 15}.
The equation of motion for skyrmions as rigid-point particles from a microscopic continuum model is derived in Ref. \cite{S-Z Lin PRB 13}.
The systematic Langevin molecular dynamics simulations of interacting skyrmions in thin films is demonstrated in Ref. \cite{Brown PRB 18} in the context of point-particle description of skyrmions.
It is shown that the interplay between Magnus force, repulsive skyrmion-skyrmion interaction
and thermal noise yields different regimes during non-equilibrium relaxation.
The spin transfer torque mechanism leads to the current-driven skyrmion  motion \cite{Schulz Nat Ph 12, Tatara PR 08}.
The experimental observation of the skyrmion Hall effect is observed
by using a current-induced spin Hall spin torque \cite{Jiang Nat Ph 17}.

The standard way to control the dynamics of skyrmions is the interaction of conduction electrons with the spin texture of the skyrmion through the
Hund's rule coupling, such that the spin of the electron is always parallel to the local magnetization in its trajectory \cite{YH Liu CPL B 15}.
This  leads to the spin transfer torque effect.
Electrons see the skyrmion as a source of flux and deviate as they pass through the skyrmion spin texture.
In the result, of the  uniform electron drift the skyrmion acquires the corresponding drift motion.
Electrons feel an additional transverse force
seeing the moving skyrmion as a source of electric field through the Faraday's law \cite{Zhang PRL 09, Tatara PRL 04}.

In addition to the electric field gradient,
the skyrmions can be controlled using magnetic field gradients in the multiferroic insulator $Cu_2OSeO_3$ \cite{Zhang Nat Comm 18}.
It is experimentally demonstrated that skyrmions rotate collectively in the field gradient.
The experimental demonstration of magnetic field gradient-driven mechanism is represented \cite{Zhang Nat Comm 18}.
The skyrmion lattice in the static magnetic field gradient is collectively rotating at lower frequencies.
The electric field control of skyrmions is a very interesting task.
It is shown that the spatial-dependent electric field, which is pointed parallel to
the direction of the average induced dipole moment of the skyrmion leads to the Hall motion of skyrmions.
The skyrmion velocity is orthogonal to the field gradient \cite{YH Liu PRB 13}.
The opportunity to control of the skyrmion motion, using the electric field gradient is demonstrated in Ref. \cite{YH Liu PRB 13}.
Coexistence of electric and magnetic fields yields the $\mathbf{E}_{eff}\times\mathbf{B}$ drift.

Skyrmions are discovered in the multiferroic
insulators \cite{Seki Sc 12, Seki PRB 12, Hurst PRB 15, Onose PRL 12}.
There is interest in development of the correlations between skyrmions dynamics and dielectric behaviors in the multiferroic materials (MFMs).
The  spiral magnetic structure in insulating materials is accompanied by finite electric dipole moment,
which is induced in the MFM, where the main role plays the pd-hybridization mechanism.
The unit cell of the MFM $Cu_2OSeO_3$ contains $16$ copper atoms and each of them is surrounded by several oxygen atoms.
The valence electron of the oxygen atom is the $p$-electron,
which has different orbital symmetry from the $d$-electron of the copper atom.
The mixing of these two orbital states leads to a redistribution of the electron cloud.
It creates the microscopic polarization on the copper-oxygen bond \cite{YH Liu CPL B 15}.
The dielectric properties of a chiral magnetic insulator $Cu_2OSeO_3$ are investigated
under various magnitudes and directions of magnetic field \cite{Seki Sc 12} and \cite{Seki PRB 12},
where authors found that each skyrmion carries the electric dipole moment.

The induced dipole moment associated with a given skyrmionic spin configuration is found in Refs. \cite{Seki Sc 12, Seki PRB 12}.
The induced electric dipole moment can be presented analytically via magnetization
$\mathbf{d}=\lambda(m^y m^z, m^x m^z, m^x m^y)$,
where $\lambda$ is a measure of the dipole moment or the magnetoelectric coupling constant \cite{YH Liu PRB 13},
and $\mathbf{m}$ is the local normalized magnetic moment.
This dipole moment is induced by the single unit cell.

The polarization of the skyrmion crystal is proportional to the dipole moment of single skyrmion $\mathbf{P}=nd\mathbf{l}$,
where $\mathbf{l}$ is the direction of the dipole moment and $\mathbf{d}=\lambda Q_DN_{sk}h/a\mathbf{l}$, where
$Q_D$ is the dipolar charge, $a$, $h$ are the linear dimension of the one unit cell and film thickness respectively. As a result skyrmions sensitive to the gradient of electric field and effect the Hall motion \cite{YH Liu PRB 13}.

We can use  a particle-like description of skyrmion and assume that the skyrmion
has a rigid internal structure in the regimes when the deformation of their internal structure is small \cite{S-Z Lin PRB 13}.
The overlap between different skyrmions is assumed to be small.

\emph{Consider a single skyrmion motion in nonuniform electric and magnetic field.}
Single skyrmion dynamics can be described by the classic Newton equation with effective mass of skyrmion
\begin{equation}\label{SK19 Newton}
m\dot{\textbf{v}}=(\textbf{d}\cdot\nabla)\textbf{E}+\frac{1}{c}[\textbf{v},(\textbf{d}\cdot\nabla)\textbf{B}]. \end{equation}
Action of the electromagnetic field on the skyrmion which has the electric dipole moment $\textbf{d}$ in accordance with data described above.
In this model, the center of mass motion of dipolar skyrmionis affected by the nonuniform electric and magnetic fields.

Consider a regime of fixed electric dipole direction during the skyrmion motion and
choose $z$-axes along the dipole moment direction $\textbf{d}=d_{0} \textbf{e}_{z}$.
We direct the magnetic field parallel to the dipole direction: $\textbf{B}=B_{0} \textbf{e}_{z}$.
Next, assume that the magnitude of the magnetic field has constant derivative in the $z$ direction $\partial B_{0}/\partial z\equiv\beta$.
If the electric field is equal to zero,
equation (\ref{SK19 Newton}) gives the following equation for the velocity of the skyrmion motion in the magnetic field
$\dot{v}_{\xi}+\imath\Omega v_{\xi}=0$,
where $v_{\xi}=v_{x}+\imath v_{y}$,
and $\Omega=d_{0}\beta/mc$ is the dipolar cyclotron frequency.
The third projection of the velocity $v_{z}$ is a constant in this regime.
It can chosen to be equal to zero $v_{z}=0$.
Hence, there is the circle motion $v_{\xi}=v_{0}e^{-\imath\Omega t}$ of the dipolar skyrmion in $x-y$ plane
with the frequency fixed by the derivative of the external magnetic field and the amplitude of the dipole moment,
where $v_{0}$ is the initial velocity.
The center of the circle orbit is motionless
or it can move in the $z$-direction creating the spiral motion of dipolar skyrmion.

Include the nonuniform part of electric field directed parallel to $y$-axis: $\textbf{E}=E_{0}(z)\textbf{e}_{y}$,
which is a function of coordinate $z$ with a constant value of the derivative $\partial E_{0}(z)/\partial z\equiv\gamma$.
The Newton equation (\ref{SK19 Newton}) reduces to the following form $\dot{v}_{\xi}+\imath\Omega v_{\xi}=\imath d_{0}\gamma/m$.
General solution appears as a superposition of solution with the zero right-hand side
and particular integral of the equation with nonzero right-hand side ($v_{\xi}=d_{0}\gamma/m\Omega$).
Hence, we obtain solution under simultaneous motion of the electric and magnetic fields
\begin{equation}\label{SK19 velocity for E and B} v_{\xi}=v_{0}e^{-\imath\Omega t}+\frac{c\gamma}{\beta}.\end{equation}
The second term in this expression gives nonzero average value at the averaging over time interval $\tau\gg\Omega^{-1}$.
The average value of found velocity $\langle \textbf{v}\rangle$ is directed parallel to the $x$ direction $\langle \textbf{v}\rangle=v_{D}\textbf{e}_{x}$ with $v_{D}=c\gamma/\beta$
($x$-direction is chosen by proper choosing of the time reference).

\begin{figure}
\includegraphics[width=8cm,angle=0]{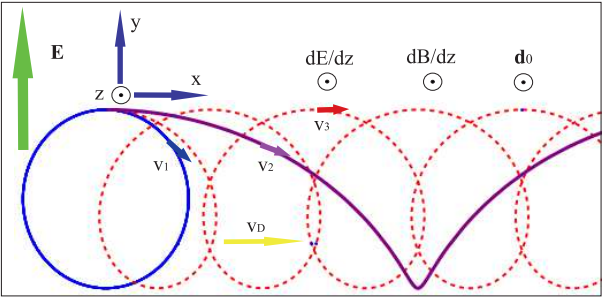}
\caption{\label{SK19 Drift Fig 01} Three regimes of trajectories are demonstrated.
Thin continuous blue circle shows regime of the zero electric field gradient.
It corresponds to rotation of dipolar skyrmions with fixed center of the orbit.
Second regime presented by dashed red line is calculated for $v_{D}=0.2 v_0$, 
where $v_0$ is the amplitude of rotation presented in equation (\ref{SK19 velocity for E and B}).
It corresponds to relatively small drift with fast rotation.
Thus, particle partially moves in direction opposite to the drift direction.
The third regime is demonstrated by the thick continuous purple line.
It is plotted for $v_{D}=1.2 v_0$.
The dipolar cyclotron frequency is the same in all regimes. 
Parameters $\textbf{v}_{1}$, $\textbf{v}_{2}$, $\textbf{v}_{3}$ illustrate the velocity of particle.}
\end{figure}

Averaging is meaningful if rotation frequency is relatively high.
It corresponds to relatively large gradients of the magnetic field.
Overwise, there are two other regimes.
Either the drift of the orbit center is small in compare with the circular motion
or there is large drift during time $t\ll\Omega^{-1}$ (some regimes are illustrated in Fig. \ref{SK19 Drift Fig 01}).
Obviously, the drift velocity should be smaller then the speed of light $v_{D}\ll c$.
It provides an estimation for the value of gradients of magnetic field $\beta\gg\gamma$.
Hence, the magnetic field gradient should be larger in compare with the electric field gradient.
Let us point out that we use the CGS units,
where the electric and magnetic fields have same physical dimensions.

The small magnetic field gradient regime can be addressed via another particular integral of the Newton equation
$v_{\xi}=\imath d_{0}\gamma t/m$ demonstrating the acceleration of dipole.
The imaginary value of $v_{\xi}$ shows that this velocity is directed in $y$-direction (in the direction of the electric field).

High magnetic field gradients provide drift of skyrmion in corresponding direction:
$\textbf{v}_{D}=2c[\partial_{z}\textbf{E},\textbf{B}]/\partial_{z}\textbf{B}^{2}$
which is perpendicular to the directions of the electric and magnetic fields.
Restore the contribution of the dipole moment in the drift velocity
\begin{equation}\label{SK19 Drift vel general}
\textbf{v}_{D}=2c\frac{[(\textbf{d}\cdot\nabla)\textbf{E},\textbf{B}]}{(\textbf{d}\cdot\nabla)\textbf{B}^{2}}. \end{equation}
It gives us the velocity of dipolar drift of skyrmion in crossed nonuniform electric and magnetic fields.

The dipolar drift motion in the magnetic field provides, let us call it, the magneto-dipolar Hall effect.
It reveals in shift of skyrmions in $x$-direction,
which gives accumulation of skyrmions on the corresponding side of the sample.

\emph{Describe collective effects in dipolar skyrmions.}
Crossed nonuniform electric and magnetic fields lead to the dipolar drift motion.
This motion can affect the collective behavior of dipolar skyrmions.
To study corresponding collective behavior use the hydrodynamic equations consisting of the continuity and Euler equations with corresponding force field:
\begin{equation}\label{SK19 continuity eq}
\partial_{t}n+\nabla\cdot(n\textbf{v})=0,\end{equation}
and
\begin{equation}\label{SK19 Euler eq}
m(\partial_{t}+\textbf{v}\cdot\nabla)\textbf{v}+\frac{\nabla p}{n}
=(\textbf{d}\cdot\nabla)\textbf{E}+\frac{1}{c}[\textbf{v},(\textbf{d}\cdot\nabla)\textbf{B}],\end{equation}
where $\textbf{d}$ is the vector field of the collective dipole moment related to the density of the electric dipole moment (the polarisability) $n\textbf{d}$,
$p$ is the thermal pressure, existing if there is unordered part of motion of skyrmions relatively to the ordered motion of collection of skyrmions.
The electric field $\textbf{E}$ and magnetic field $\textbf{B}$ are the external fields
if the interaction between skyrmions is neglected.
The account of the dipole-dipole interaction between dipolar skyrmions generalize the electric field
to the superposition of the external field $\textbf{E}_{ext}$ and the electric field caused by the dipoles $\textbf{E}_{int}$.
The interacting electric field obeys the Poisson equation:
\begin{equation}\label{SK19 Poisson eq} \nabla\cdot\textbf{E}_{int}=-4\pi\textbf{d}\cdot\nabla n.\end{equation}
The Poisson equation is written for the dipole moments with fixed direction.
The polarization is created by corresponding background electric field $\textbf{E}_{b}=E_{b}\textbf{e}_{z}$.
Presented hydrodynamic equations allows to study waves and instabilities of skyrmions in crossed nonuniform electric and magnetic fields.

\emph{First focus on instabilities of ideal skyrmion gas.}
Consider the electric field gradient affecting the collective behavior of skyrmions.
The external electric field creates nonzero equilibrium force field.
Choose the electric field to be directed in $y$-direction with the gradient directed in $z$-direction:
$\textbf{E}=E_{0}(z)\textbf{e}_{y}$,
which is a nonuniform part of full external electric field
$\textbf{E}_{ext}=E_{b}\textbf{e}_{z}+E_{0}\textbf{e}_{y}$,
where $E_{b}\gg E_{0}$.
Hence, it is necessary to create an equilibrium pressure gradient caused
by the temperature gradient to create a macroscopically motionless equilibrium with uniform concentration.
Find the following equilibrium condition $\nabla p_{0}=n_{0}d_{0}(\partial E_{0}/\partial z)$,
where $n_{0}$ is the constant equilibrium concentration,
$d_{0}$ is the equilibrium constant electric dipole moment,
and $p_{0}(T)$ is the equilibrium pressure as a function of nonuniform temperature $T(y)$.
Consider plane wave perturbations of the described equilibrium state obtain
the following simplification of the hydrodynamic equation appearing in the linear regime
$\partial_{t}\delta n +n_{0}\partial_{y}\delta v_{y}=0$,
$mn_{0}\partial_{t}\delta v_{y}=d_{0}\gamma\delta n$,
where $\partial E_{y}/\partial z\equiv\gamma$.
Use the following ansatz
$\delta n=N e^{-\imath\omega t+\imath k_{y}y}$,
and $\delta v_{y}=U_{y} e^{-\imath\omega t+\imath k_{y}y}$,
where $\omega$ is the frequency of the wave,
$k_{y}$ is the projection of the wave vector on the $y$-direction,
$N$ and $U_{y}$ are constant amplitudes of perturbations,
to reduce the linear differential equations to the algebraic equations.
They results in the following dispersion dependence
\begin{equation}\label{SK19 el field inst} \omega=\frac{1+\imath}{\sqrt{2}}\sqrt{\frac{k_{y}\gamma d_{0}}{m}}.\end{equation}

Let us introduce the isotropic characteristic frequency
of the electric field caused instability for the further references $\Delta_{0}\equiv\sqrt{\gamma d_{0}k/m}$.

Start the small amplitude perturbation analysis with a simplified regime of zero electric field gradient contribution.
In this regime the system is characterized by the equilibrium concentration $n_{0}$, equilibrium pressure $p_{0}$, equilibrium collective dipole moment $\textbf{d}_{0}=d_{0} \textbf{e}_{z}$, and the spatial derivative of the external magnetic field $\partial B_{z}/\partial z=\beta$.
For small perturbations represent physical parameters as follows
$n=n_{0}+\delta n$, $\textbf{v}=0+\delta \textbf{v}$, $p=p_{0}+\delta p$,
with $\delta p=mU^{2}\delta n$,
where $U^{2}$ is the mean-square thermal velocity.
It is assumed that perturbations are proportional to $e^{-\imath\omega t+\imath \textbf{k}\textbf{r}}$,
where $\textbf{k}=\{ k_{x}, k_{y}, k_{z}\}$.

Linearized hydrodynamic equations lead to the anisotropic spectrum of sound waves:
\begin{equation}\label{SK19 anisotr sound}
\omega^{2}=\frac{1}{2}\biggl[\Omega^{2}+k^{2}U^{2}
\pm\sqrt{\Omega^{4}+2\Omega^{2}U^{2}(k_{\perp}^{2}-k_{z}^{2})+k^{4}U^{4}}\biggr],\end{equation}
where the anisotropy is caused by the dipolar cyclotron frequency $\Omega$, and $k_{\perp}^{2}=k_{x}^{2}+k_{y}^{2}$.
It gives the following simplified result for waves propagating perpendicular to the direction of the dipole moment ($k_{z}=0$)
$\omega^{2}=\omega_{\perp}^{2}$,
where $\omega_{\perp}^{2}=\Omega^{2}+k_{\perp}^{2}U^{2}$.
There is single wave in this regime.
It is a sound-like wave spectrum shifted by the dipolar cyclotron frequency,
which makes the gap in the spectrum.
The regime of propagation parallel to the direction of the dipole moment ($k_{\perp}=0$) gives two solutions:
the rotation with fixed dipolar-cyclotron frequency $\omega^{2}_{+}=\Omega^{2}$
and the sound wave $\omega^{2}_{-}=k_{z}^{2}U^{2}$.

Assuming that the dipolar cyclotron frequency dominates
we can simplify equation (\ref{SK19 anisotr sound}) for arbitrary wave vector direction $\textbf{k}$:
$\omega^{2}_{+}=\Omega^{2}+k_{\perp}^{2}U^{2}$,
and $\omega^{2}_{-}=k_{z}^{2}U^{2}$.
Hence, the second wave is a gapless sound wave propagating parallel to the electric dipole moment.

Include the external electric field directed parallel to $y$ direction and changing its magnitude along $z$ direction
include in our analysis the spatial derivative of the external electric field $\gamma$.
For waves propagating in the plane $x-y$, perpendicular to the direction of electric dipole moment, find corresponding linearized hydrodynamic equations
$\delta n=n_{0}(k_{x}\delta v_{x}+k_{y}\delta v_{y})/\omega$,
\begin{equation}\label{SK19 Euler lin x} \omega \delta v_{x}- k_{x} U^{2}\frac{\delta n}{n_{0}}=\imath \Omega\delta v_{y}, \end{equation}
and
\begin{equation}\label{SK19 Euler lin y} \omega \delta v_{y}- k_{y} U^{2}\frac{\delta n}{n_{0}}
=\frac{\imath\gamma d_{0}}{mn_{0}} \delta n -\imath \Omega\delta v_{x}.\end{equation}
This set of equations gives the following dispersion equation
\begin{equation}\label{SK19 dispersion eq for anisotr sound in plane}
\omega^{2}-\omega_{\perp}^{2}+\biggl(\frac{\Omega}{\omega}k_{x}-\imath k_{y}\biggr)\frac{\gamma d_{0}}{m}=0.\end{equation}
The multiplier $1/\omega$ in the last term of equation (\ref{SK19 dispersion eq for anisotr sound in plane})
increases the degree of the dispersion equation.
Assuming that the contribution of the electric field is relatively small
(as it is correspond to the drift motion condition)
solve this equation by the iteration method and obtain
\begin{equation}\label{SK19 frequency for anisotr sound in plane}
\omega^{2}=\omega_{\perp}^{2}
+\biggl(\imath k_{y}-\frac{\Omega}{\omega_{\perp}}k_{x}\biggr)\frac{\gamma d_{0}}{m}.\end{equation}
Solution of equation (\ref{SK19 frequency for anisotr sound in plane}) demonstrates small in-plane anisotropy of the sound frequency
\begin{equation}\label{SK19 frequency Re for anisotr sound in plane}
\omega_{R}^{2}=\Omega^{2}+k_{\perp}^{2}U^{2}
-\frac{\Omega}{\omega_{\perp}}k_{x}\frac{\gamma d_{0}}{m},\end{equation}
and presence of instability with increment
\begin{equation}\label{SK19 frequency Im for anisotr sound in plane}
\delta= k_{y}\frac{\gamma d_{0}}{m\omega_{R}}\approx k_{y}\frac{\gamma d_{0}}{m\Omega},\end{equation}
where $\omega=\omega_{R}+\imath\delta$.
The increment (\ref{SK19 frequency Im for anisotr sound in plane}) shows prominent anisotropy.
The approximate relation in equation (\ref{SK19 frequency for anisotr sound in plane}) is made for small pressure effects.
Solution (\ref{SK19 frequency for anisotr sound in plane}) is found in the approximation of small instability.
Hence, it shows that as small time scale the magnetic field change the instability.
While the following growth of instability leads to a transition thought intermediate regime to the electric field instability demonstrated above (\ref{SK19 el field inst}).
The following growth of perturbation amplitude in accordance with the electric field instability carries described system to the nonlinear regime.

The increment (\ref{SK19 frequency Im for anisotr sound in plane}) can be rewritten via the characteristic frequency $\Delta_{0}$: $(k_{y}/k)(\Delta_{0}^{2}/\Omega)$.

If $k_{y}=0$ equation (\ref{SK19 frequency for anisotr sound in plane}) shows a possibility for another instability.
It is related to the last term
which proportional to the gradient of the electric field $\gamma$
and the gradient of magnetic field hidden in the dipolar cyclotron frequency $\Omega$ simultaneously.
Dropping the thermal effects $\sim U^{2}$ (hence $\omega_{\perp}\approx\Omega$)
find simplification of equation (\ref{SK19 frequency for anisotr sound in plane}) for $k_{y}=0$:
$\omega^{2}=\Omega^{2}- \gamma d_{0}k_{x}/m$.
The last term does not contain any trace of the magnetic field,
but its appearance is related to the presence of the magnetic field.
Hence, there is critical wave vector $k_{0x}=(\beta/\gamma)\Omega/c$,
where the instability appears.
The dipolar drift corresponds to $\beta/\gamma\gg1$.

In this regime, the structure of term causing instability is similar to described above.
Hence, it can be presented as $-k_{x}\Delta_{0}^{2}/k$.

Electric field gradient $\gamma=3\cdot10^{3}$ CGS units/cm ($\gamma=10^{10}$ $V/m^2$) \cite{YH Liu PRB 13}
and magnetic field gradient $\beta=10$ $G/cm$ (0.1 $mT/mm$) \cite{Zhang Nat Comm 18}
and $\beta=1.8\cdot 10^{7}$ $G/cm$ ($\beta=1.8\cdot 10^{2}$ $T/mm$) \cite{Wang NJP 17}
are characteristic examples of the field gradients applied for the manipulation of skyrmions.
However, the relation between $\gamma$ and $\beta$ is opposite to the condition required for the magneto-dipolar drift of dipolar skyrmions.
To reach required condition one needs to generate $\beta=10^{7}$ $G/cm$ ($10^{2}$ $T/mm$) like in Ref. \cite{Wang NJP 17},
at $\gamma=1\div10$ CGS units/cm ($\gamma=10^6\div10^{7}$ $V/m^2$).
It gives the following drift velocity $v_{D}=3\cdot10^{3}$ $cm/s$ ($\gamma=1$ CGS units/cm, $\beta=10^{7}$ $G/cm$).

Including the characteristic value of the electric dipole moment generate MFMs
$d_{0}=3\cdot 10^{-21}$ CGS units ($d_{0}=10^{-32}$ $C\cdot m$) \cite{YH Liu PRB 13},
and effective mass of skyrmions $m=3\times10^{-21}$ g ($3\cdot10^{-24}$ kg) \cite{Yang OEx 18}
estimate the dipolar cyclotron frequency for $\beta=10^{7}$ $G/cm$.
Hence, we find $\Omega=10^{-3}$ $s^{-1}$.
It is rather small frequency.
Therefore, there is no rotation in this regime while particle moves in direction perpendicular to the electric field.
In $y$-direction the skyrmion accelerates up to $0.3$ $cm/s$ during $1$ s at $\gamma=1$ CGS units/cm.
Hence, it moves mostly in $x$-direction.

The critical wave vector $k_{0x}\approx0$ is negligible small.
Hence, the magneto-dipolar drift instability happens for all wave vectors.

\emph{Role of dipole-dipole interaction in instabilities of weakly interacting skyrmion gas.}
It is also essential to estimate the role of inter-skyrmions interactions,
which have the dipole-dipole nature for the MFM.
The dipole-dipole interaction gives a generalization of equations (\ref{SK19 Euler lin x}), (\ref{SK19 Euler lin y}),
which includes the nontrivial third projection of the wave vector together with the z-projection of the velocity.
Moreover, the following term rises on the right-hand side of linearized Euler equation $n_{0}d_{0}k_{z}\textbf{k}\delta\varphi_{int}$,
where $\delta\varphi_{int}$ is the perturbations of scalar potential of electric field obeying the linearized Poisson equation
$k^{2}\delta\varphi_{int}=4\pi d_{0}\imath k_{z} \delta n$.
Described equations give the dispersion equation
which appears in the following form
$$\omega^{4}-\omega^{2}(\Omega^{2}+k^{2}V^{2}+\imath\gamma d_{0} k_{y})$$
\begin{equation}\label{SK19 disp eq at int}
+\Omega\gamma d_{0} k_{x}\omega+\Omega^{2}k_{z}^{2}V^{2}=0,\end{equation}
where $k^{2}=k_{\perp}^{2}+k_{z}^{2}$ and $V^{2}=U^{2}+\Lambda^{2}k_{z}^{2}/k^{2}$,
with the characteristic frequency for the dipole-dipole interaction
$\Lambda^{2}=4\pi d_{0}^{2}k^2/m$.
Numerical estimation of characteristic frequency of dipole-dipole interaction for
$d_{0}=3\cdot 10^{-21}$ CGS units and $m=3\cdot 10^{-21}$ $g$ gives $\Lambda\approx3\cdot 10^{-10}k$ $s^{-1}$.
For the wave length of order of 1 $\mu m$ we have $\Lambda\approx3\cdot 10^{-6}$ $s^{-1}$ ($\Lambda>\Omega$).

If the contribution of the electric field is small $\gamma\rightarrow0$,
find stable solutions of dispersion equation (\ref{SK19 disp eq at int}):
\begin{equation}\label{SK19 freq at int 0}
\omega_{0}^{2}=\frac{1}{2}\biggl[w^{2}\pm\sqrt{w^{4}-4\Omega^{2}k_{z}^{2}(U^{2}+\Lambda^{2}\cos^{2}\theta)}\biggr], \end{equation}
where $w^{2}\equiv\Omega^{2}+k^{2}(U^{2}+\Lambda^{2}\cos^{2}\theta)$, and $k_{z}=k\cos\theta$.
This is a generalization of equation (\ref{SK19 anisotr sound}).

The dipole-dipole interaction gives a positive nonisotropic shift of the thermal velocity square $U^{2}$.
Hence, the shift is maximal if the wave propagate in the direction of the electric dipole moment.
There is no dipole-dipole interaction contribution for the perpendicular direction.

Comparing this result with equations
(\ref{SK19 frequency Re for anisotr sound in plane}) and (\ref{SK19 frequency Im for anisotr sound in plane})
we conclude that the dipole-dipole interaction decreases the instability increment for the nonzero electric field gradient $\gamma\neq0$.

Moreover, the dipole-dipole interaction increases the value of the critical wave vector $k_{0x}$
for the dipolar drift instability increases for nonzero $k_{z}$.
What leads to small stabilization of the system.
In equation (\ref{SK19 disp eq at int}) the dipolar drift instability is presented by the term linear on frequency $\omega$.

\emph{Conclusions.}--
The experimental data obtained for MFMs show
that the skyrmions in MFMs have the electric dipole moment.
We have considered the dynamics of single dipolar skyrmions and gas of weakly interacting dipolar skyrmions.
We have applied the Newton and hydrodynamic models for the point-like interacting skyrmions
in the external electromagnetic fields to study the drift motion of dipolar skyrmions.
We have considered the influence of the nonuniform magnetic field on the moving dipole moment.

We have found new way to control of skyrmion motion using magnetic field gradient.
The large enough magnetic field gradient applied together with the electric field gradient generate the dipolar drift motion.
The velocity of dipolar drift of skyrmion in the crossed nonuniform electric and magnetic fields has been derived.
The drift velocity is perpendicular to the directions of the nonuniform electric field
and to the direction of magnetic field.
As a result we have obtained the magneto-dipolar Hall effect in crossed nonuniform electric and magnetic fields.
It provides an extra mechanism for control of the skyrmion motion and distribution of collection of skyrmions.

Collective dynamics of skyrmions has been addressed through the hydrodynamic model.
We have found three regimes of collective instabilities in systems of dipolar skyrmions in the MFMs.
One corresponds to acceleration of dipoles by the nonuniform electric field.
Other two regime are related to simultaneously acting electric and magnetic fields.
One of them change the small time regime of instability happening in the direction of electric field.
The another one is the instability happening in the direction of the drift motion, i.e. the dipolar drift instability.

\emph{Acknowledgments.}--
The study of M. Trukhanova was performed by a grant of the Russian Science Foundation (Project N 19-72-00017).

\end{document}